\documentclass[a4paper,10pt]{article}
\usepackage{
times,
amsmath,amssymb,epsfig,graphics,graphicx}

\newcommand{\be}{\begin{eqnarray}}
\newcommand{\ee}{\end{eqnarray}}

\begin{document}
%\date{\today }

\begin{center}
{ \Large \bf Symmetry and degeneracy of the curved Coulomb potential on the
$S^3$ ball     }
\end{center}

\begin{center}
A.\ Pallares-Rivera and M.\ Kirchbach\\
Instituto de F\'{\i}sica, \\
       Universidad Aut\'onoma de San Luis Potos\'{\i},\\
       Av. Manuel Nava 6, San Luis Potos\'{\i}, S.L.P. 78290, M\'exico
\end{center}

\begin{flushleft}
{\bf Abstract}:
The ``curved'' Coulomb potential on the $S^3$ ball, 
whose isometry group is SO(4), takes the
form of a cotangent function, and  when added to
the four-dimensional squared angular momentum operator, 
one of the so(4) Casimir invariants, a Hamiltonian is obtained  
which describes a  perturbance of the free geodesic motion
that results peculiar in several aspects.
The spectrum  of such a  motion has been studied on various occasions 
and is known to carry
unexpectedly  so(4) degeneracy patterns despite the
non-commutativity of the perturbance with the  Casimir operator. 
We here suggest an explanation for this behavior
in designing a set of operators which close the so(4) algebra
and whose  Casimir invariant coincides with the Hamiltonian of the perturbed
motion at the level of the eigenvalue problem. The above operators are related
to the canonical geometric SO(4) generators on $S^3$ by a
non-unitary similarity
transformation of the scaling type.
In this fashion, we identify
a complementary  option to the deformed dynamical so(4) Higgs  algebra
constructed in terms of the components of the ordinary angular momentum and
a related  Runge-Lenz vector.
\end{flushleft}

\begin{flushleft}
\end{flushleft}

\section{Introduction}
The three-dimensional hyper-spherical surface, $S^3$,
embedded in a four-dimensional Euclidean space, $E_4$, 
\begin{equation}
S^3:\quad x_4^2+x_1^2+x_2^2+x_3^2=R^2,
\label{I1}
\end{equation}
is among the most important geometries in a variety of theoretical
physics problems. In first place this curved surface, whose isometry 
group is SO(4), represents the position space
in the celebrated  Einstein's closed Universe and provides one of
the major templates in gravitational studies.
At the same time, it forms part of the compactified Minkowski
spacetime as it emerges upon conformal compactifications in
supersymmetric field  theories. Further important aspects of the
$S^3$ ball concern its relevance in the effective description of
many-body phenomena  such as coherent states, 
fluid-dynamics, \cite{PRA}-\cite{PRL}, polymer chains \cite{poly}, 
Brownian motion \cite{Brown}, and quantum dots \cite{qtm_dots},  
all reasons which make studies of physics on  $S^3$ relevant.

The free geodesic motion of a scalar particle  
on $S^3$ is described by means of the eigenvalue problem of the 
squared four-dimensional angular momentum, ${\mathcal K}$, one of the
Casimir invariants of the so(4) isometry algebra.
We here focus on a perturbance of this motion by a $\cot \chi$ 
function of the second polar angle, $\chi$,
parametrizing $S^3$,  which  is interesting 
in so far as it conserves the SO(4) degeneracy patterns in the spectrum of
the free geodesic motion, despite its non-commutativity with ${\mathcal K}$.
This property of the cotangent function
is quite remarkable indeed and valid not only on $S^3$ but also on $S^2$,
and in any higher dimensions.
To be specific, also on $S^2$ the cotangent potential does not remove
the $(2 \ell+1)$-fold degeneracy in
the spectrum  of  the free spherical
rigid rotator, despite its non-commutativity
with ${\mathbf L}^2$ \cite{Higgs}, \cite{MolPhys}.
For the explanation of the above degeneracy phenomena,
Higgs and Leemon  designed in \cite{Higgs},
\cite{Leemon} algebras composed by the components
of the  angular momentum operators, $L_i$, in the respective external 
spaces under consideration, and a related   
Runge-Lenz vector, $R_j$, designed  in analogy to the flat-space one,
as known from the H atom problem.
This strategy has been successful in explaining the degeneracy phenomena
under discussion, despite that $L_j$ and $R_i$
cease to close the respective so(3)/so(4) algebras. Instead, they close
algebras which appear as deformations of the respective
isometry  algebras by terms
cubic in the momenta (see also \cite{Quesne} for further details). \\

\noindent
We here instead  construct a closed
so(4) algebra for the cotangent perturbed  motion on
$S^3$, which we obtain as a non-unitary similarity transformation of the
geometric so(4)  algebra of the free geodesic motion.
Such becomes possible at the level of the  
$\left( {\mathcal K}-2b\cot\chi\right)$ and ${\mathcal K}$
eigenvalue problems.
We build up a  matrix similarity transformation which
connects the ${\mathcal K}$ and $({\mathcal K}- 2b\cot\chi )$
carrier spaces,  and which happens to be of the dilation (scaling) type.
The possibility for constructing such a transformation on $S^2$ has been 
indicated in work prior to this \cite{MolPhys} and will not be 
considered here.\\

We entirely focus on the so(4) case which is  
special by the fact that on $S^3$ the
cotangent function  solves the homogeneous Laplace-Beltrami
equation, and is a so called harmonic function there. As such,
it can be treated along the line
of potential theory and the resulting electrodynamics would be Maxwellian,
something which does not occur in any other dimension.

\noindent
The paper is structured as follows.
In the next section we present the solutions of the
cotangent-perturbed geodesic motion
on $S^3$ in terms of real Romanovski polynomials (reviewed in \cite{raposo})
and decompose these solutions  in the basis of exponentially damped
hyper-spherical harmonics. From there we deduce the
non-unitary transformation which takes the so(4) isometry algebra to
a realization that  now 
describes the cotangent-hindered motion.
Working at the level of the
eigenvalue problems brings  as an advantage simplifications  by virtue
of certain type of  recurrence relations among Gegenbauer polynomials.
The paper closes with concise conclusions.

\section{Particle motion on the  $S^3$ ball}

\subsection{The free geodesic motion}
In polar coordinates, the  three-sphere $S^3$ is parametrized 
by the azimuthal angle $\varphi$, and the two
polar angles $\theta$, and $\chi$,  according to,
\begin{eqnarray}
S^3:&\quad& x_4^2+x_1^2+x_2^2+x_3^2=R^2, \quad x_4=R\cos\chi, \quad
r\equiv \sqrt{x_1^2+x_2^2+x_3^2}=R\sin\chi, \nonumber\\
&\quad& x_3=r\cos\theta, \quad x_1=r\sin\theta \cos \varphi, \quad
x_2=r\sin\theta \sin\varphi\, ,
\label{I2}
\end{eqnarray}
where $R$ will be treated as a constant and will be set equal to one
for simplicity. The  Casimir operator,
${\mathcal K}$,  of the so(4) algebra
is associated with the squared four-dimensional angular momentum, and is
given by \cite{Engelfield},
\begin{eqnarray}
{\mathcal K}&=&{\mathbf L}^2 +{\mathbf N}^2,
\quad L_i=i\epsilon_{ijk}x_j\frac{\partial}{\partial x_k},
\quad N_j= -ix_j\frac{\partial}{\partial x_4}+
ix_4\frac{\partial}{\partial x_j},
\nonumber\\
{\mathbf N}^2&=&
-\frac{1}{\sin\chi}\frac{\partial }{\partial \chi }\sin \chi
\frac{\partial }{\partial \chi }  +\cot^2\chi \, {\mathbf L}^2,
\label{I3}
\end{eqnarray}
where the components $L_i$, and $N_j$  of the respective
angular momentum, ${\mathbf L}$, and the Euclidean boost operator,
${\mathbf N}$, have the property to close the so(4) algebra.
The  quantum-mechanical ${\mathcal K}$ eigenvalue problem is well known 
and given by,
\begin{eqnarray}
\frac{\hbar^2}{2M}{\mathcal K} \, Y_{K \ell m}(\chi, \theta, \varphi)&=&
\frac{\hbar^2}{2M}
\left[\left(K+1\right)^2-1\right]Y_{K \ell m}(\chi,\theta, \varphi),
\nonumber\\
Y_{K \ell m}(\chi, \theta,\varphi)&=&\
\sin^{\ell}\chi {\mathcal G}_{K-\ell}^{\ell+1}(\cos \chi)Y_{\ell}^m(\theta, \varphi),
\label{I4}
\end{eqnarray}
where the constant $K$ determines the value of the four-dimensional
angular momentum in the $(K+1)^2$-dimensional
representation space of the hyper-spherical harmonics,
$Y_{K \ell m}(\chi,\theta, \varphi)$,
${\mathcal G}_{n=K-l}^{\ell+1}(\cos\chi)$ denote the Gegenbauer polynomials, and
$Y_{\ell}^m(\theta, \varphi)$  are the standard three-dimensional
spherical harmonics. We have chosen to work in dimensionless units,
setting $\hbar =1$, and $2M=1$, with $M$ standing for the mass of the 
particle  under consideration.
It is obvious that the spectrum of the free geodesic motion on $S^3$
in (\ref{I4}) is characterized by
a $(K+1)^2$-fold degeneracy of the states, as it should be given the
fact that $SO(4)$ is the isometry group of the three-ball.
The immediate conspicuous analogy that  comes to ones mind is that
modulo the level spacings, the degeneracy patterns of the
free geodesic motion on $S^3$ are same as   
those appearing in the inverse distance potential problem
which shapes the H atom spectrum. Yet,  
the reasons for the two phenomena  are to some extent different, indeed.
The common denominator of both degeneracy patterns is an underlying
so(4) symmetry algebra of the associated Hamiltonians.
The difference lies in the qualitatively distinct
realizations of this very algebra in the respective cases.
While the so(4) algebra of the free geodesic motion on $S^3$ is purely
geometric in the sense that its elements, $L_i$, and $N_j$ in (\ref{I3}) 
exclusively depend on the
position on the curved surface under consideration,
the so(4) algebra in the inverse distance potential problem is
dynamical as its elements need to be defined over the
full phase space.  Indeed, the so(4) algebra of the Coulomb potential
contains besides the components
$L_1$, $L_2$, and $L_3$, of ordinary angular momentum,  the three
components of the Runge-Lenz vector, $R_1$, $R_2$, and $R_3$ \cite{Engelfield}.
Below, this concept will acquire special importance in the investigation of
the perturbed problem.

Combining eqs.~(\ref{I3}) and (\ref{I4}), the eigenvalue problem of
the squared four-dimensional  angular momentum operator becomes,

\begin{eqnarray}
\left[
-{1 \over \sin^2 \chi }{\partial \over \partial \chi}
\sin^2 \chi {\partial \over \partial \chi}
+{\pmb{L}^2 \over \sin^2 \chi} \right] Y_{K{\ell}m}(\chi, \theta, \varphi)
&=&  K(K+2)\, Y_{K{\ell}m}(\chi, \theta, \varphi),
\label{S1_1}
\end{eqnarray}
with the hyper-spherical harmonics, $Y_{K{\ell}m}(\chi,\theta,\varphi)$,
being defined in (\ref{I4}) above.
In what follows we introduce the short-hand,
\begin{eqnarray}\label{S1_2}
\mathcal{S}^{\ell}_{K}(\chi)&=&
\sin^{\ell} \chi \; \mathcal{G}^{\ell+1}_{K-\ell} (\cos \chi),
\label{short_hand_S}
\end{eqnarray}
in terms of which the hyper-spherical harmonics equivalently rewrite to,
\begin{eqnarray}
Y_{K{\ell}m}(\chi,\theta,\varphi)
&=& {\mathcal S}^{\ell}_K (\chi)Y^{m}_{\ell}(\theta,\varphi),
\quad K \in [0, \infty ),\quad \ell \in [0,K],\quad  m\in[-\ell, \ell].
\label{S1_3}
\end{eqnarray}

\subsection{The  ``curved'' Coulomb potential problem on $S^3$}
For the sake of self-sufficiency of the presentation and
fixing notations we here briefly review the exact solutions of the
``curved'' Coulomb (-like)  potential on
$S^3$, given by,
\begin{equation}
\left({\mathcal K}-2b\cot\chi \right)\Psi_{K{\widetilde \ell}{ \widetilde m}}
(\chi, \theta,\varphi)
=\epsilon \Psi_{K{\widetilde \ell}{\widetilde  m}}(\chi, \theta,\varphi).
\label{I1_1}
\end{equation}
before going to the heart of our study in the next subsection,
namely, to their expansions 
in the basis of the hyper-spherical harmonics.
In (\ref{I1_1}), $\epsilon$ is the energy, $E$, here in dimensionless units,
$\epsilon =2ME/\hbar^2$, and $b$ defines the (equally dimensionless)
strength of the cotangent potential.
We furthermore admitted in (\ref{I1_1}) for the possibility that
${\widetilde m}$ and ${\widetilde \ell}$ may not necessarily
be same as $m$ and $\ell$ in  (\ref{I4}), (\ref{S1_1}).

Upon changing variable in (\ref{I1_1}) to,
\begin{equation}
\Psi_{K {\widetilde \ell}{ \widetilde m}}(\chi,\theta,\varphi)=
\frac{U_K^{\widetilde \ell }(\chi)}{\sin\chi }
Y_{{\widetilde \ell}}^{\widetilde m}(\theta, \varphi),
\label{I5}
\end{equation}
equation (\ref{I1_1}) transforms into the well known one-dimensional
Schr\"odinger equation with the so called trigonometric Rosen-Morse potential
\cite{Dutt},
${\mathcal V}_{\mbox{RM}}(\chi )$,
in reality introduced  by Schr\"odinger \cite{Schr40}, \cite{Schr41}
as a ``curved''  Coulomb-, better,  Coulomb-like potential \cite{Barut},
\begin{eqnarray}
-\frac{{\mathrm d}^2U_K^{\widetilde \ell }(\chi)}{{\mathrm d}\chi^2}
&+&{\mathcal V}_{\mbox{RM}}(\chi)U_K^{\widetilde \ell
}(\chi)-(\epsilon +1)U_K^{\widetilde \ell }(\chi )=0,\nonumber\\
{\mathcal V}_{\mbox{RM}}(\chi )&=&-2b\cot\chi +\frac{ \ell (\ell+1)}{\sin^2\chi},
\label{I6}
\end{eqnarray}
in depending on the choice for the $b$ value.
The latter equation is exactly solvable because it can be reduced
to the hyper-geometric differential equation \cite{Dutt}, and the spectrum is such that the
cotangent perturbance does not remove the $(K+1)^2$-fold 
degeneracy of the free geodesic
motion, as visible from the expression for the energy,
\begin{equation}
\epsilon_K +1 = (K+1)^2 -\frac{b^2}{(K+1)^2}.
\label{I7}
\end{equation}
 The dependence of the excitation energies in eq.~(\ref{I7})
on the strength $b$ of the cotangent perturbance is displayed in Fig.~1. 
\begin{figure}
\resizebox{0.80\textwidth}{7.5cm}
{\includegraphics{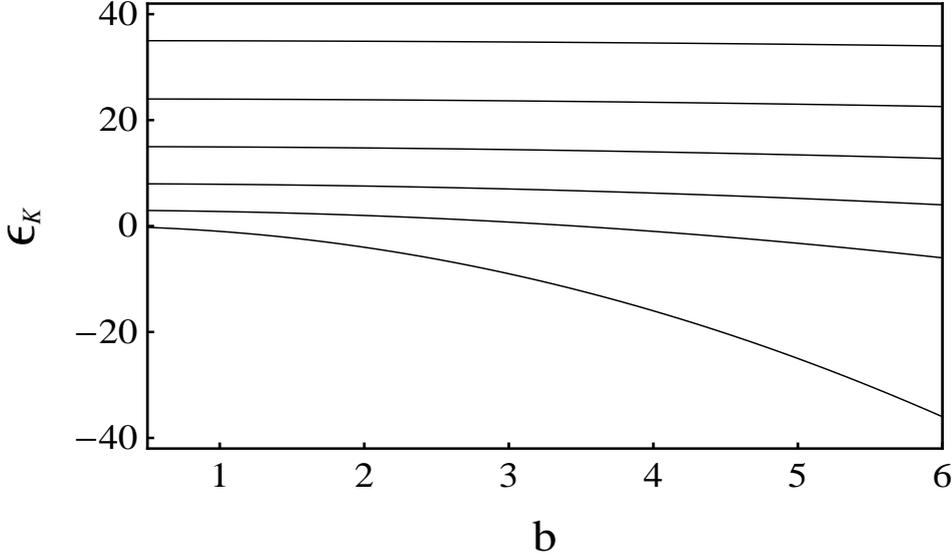}}
\caption{The dependence of the energies, $\epsilon_K$, in (\ref{I7}), of the
levels within  the ``curved'' Coulomb potential  
on the strength $b$ of the $\cot\chi $ perturbance. The figure 
shows that the cotangent interaction mainly affects the
gap between the ground state and its first excitations by increasing it.
For moderate values of the potential strength around $b\sim 1$,
 the higher lying states 
practically remain unaltered by the perturbance due to the rapid 
flattening of the exponential factor in (\ref{S1_5}) with the increase of
$K$.
The exponential damping factor furthermore
reduces  the de-excitation probabilities to the ground state,
the most affected being  the first excitation.
\label{energy}}
\end{figure}

Among others, the solutions in (\ref{I5}) have been independently reproduced
also  in \cite{KC_10} in terms of non-classical Romanovski polynomials,
$R_n^{\alpha , \beta}(\cot \chi )$, as
\begin{eqnarray}\label{S1_5}
\Psi_{K\, \widetilde{\ell}\, \widetilde{m}}(\chi,\theta,\varphi)
&=& e^{-\frac{\alpha_K \chi}{ 2}} \sin^K\chi\,
R^{\alpha_K, \beta_K}_{K -\widetilde{\ell}}(\cot \chi)\,
Y^{\widetilde{m}}_{\widetilde{\ell}}(\theta,\varphi),\nonumber\\
\beta_K &=& -(n + \widetilde{\ell})= -K,
\quad \alpha_K = \frac{2b}{K+1}.
\end{eqnarray}
The Romanovski polynomials (reviewed in ref.~\cite{raposo})
satisfy the following differential hyper-geometric equation:
\begin{equation}
(1+x^2)\frac{{\mathrm d}^2R_n^{\alpha, \beta}}{{\mathrm d} x^2}
+2\left(\frac{\alpha }{2} +\beta x
\right)\frac{{\mathrm d}R_n^{\alpha, \beta}}{{\mathrm d}x}
-n(2\beta +n-1)R_n^{\alpha, \beta}=0.
\end{equation}
They are obtained from the following  weight function,
\begin{equation}
\omega ^{\alpha, \beta}(x)=(1+x^2)^{\beta -1}\exp(-\alpha \cot^{-1}x),
\end{equation}
by means of the Rodrigues formula,
\begin{equation}
R^{\alpha, \beta}_n(x)=\frac{1}{\omega^{\alpha, \beta }(x)}
\frac{{\mathrm d}^n}{{\mathrm d}x^n}
\left[ (1+x^2)^n\omega^{\alpha, \beta}(x)\right].
\end{equation}
Upon introducing the short-hand,
\begin{eqnarray}
\psi^{\,\widetilde{\ell}}_{\,K}\, (\chi) &=&
\sin^K\chi R^{\alpha_K, \beta_K}_{K -\widetilde{\ell}}(\cot \chi),
\label{S1_6}
\end{eqnarray}
one arrives at the final form of the solutions to eq.~(\ref{I1_1}),
\begin{eqnarray}
\Psi_{K\, \widetilde{\ell}\, \widetilde{m}}(\chi,\theta,\varphi)
&=& e^{-\frac{\alpha_K \chi}{  2}}\, \psi^{\,\widetilde{\ell}}_{\,K}\, (\chi)\,
Y^{\widetilde{m}}_{\widetilde{\ell}}(\theta,\varphi).
\label{S1_7}
\end{eqnarray}

\subsection{The perturbed motion in terms  of damped 
hyper-spherical harmonics }
The goal of the present section
is to find finite decompositions of the exact solutions of
the ``curved''Coulomb(-like) potential on $S^3$ in the basis of the canonical
hyper-spherical harmonics describing the free geodesic motion.
We begin with seeking to relate  the $\chi$ dependent parts,
$\psi_K^{\widetilde{\ell}}(\chi)$, and ${\mathcal S}_K^{\ell}(\chi)$, of the
wave functions of the respective perturbed, and free motions
in (\ref{S1_6}), and  (\ref{S1_2}), as

\begin{eqnarray}
\psi^{\,\widetilde{\ell}}_{\,K}\, (\chi)
&=& \sum_{\ell = \widetilde{\ell}}^{K} C_{\ell}\, \mathcal{S}^{\ell}_{K}(\chi)=
\sum_{\ell = \widetilde{\ell}}^{K} C_{\ell}\,
\sin^{\ell} \chi \, \mathcal{G}^{\ell + 1}_{K-\ell} (\cos \chi).
\label{S1_8}
\end{eqnarray}
The latter equation in fact represents a new ansatz for constructing 
$\psi_K^{\widetilde \ell}(\chi)$  in (\ref{S1_6}),  and it is indeed possible
to write down series of conditions which  fix the constants $C_{\ell}$.
However, one can equally well  take advantage of already
knowing $\psi_K^{\widetilde \ell}(\chi)$, and encounter the expansion coefficients
in (\ref{S1_8})  using the orthogonality properties of the hyper-spherical 
harmonics. Both ways are eligible. We here opt for the second one, and
 encounter  the expansions given in Table 1 below.
Substituting them in eq.~(\ref{S1_7}), and making use of the identity,
\begin{equation}
{\mathcal S}_K^{\ell}(\chi )\equiv \frac{e^{-im\varphi}}{P_{\ell}^{m}(\cos \theta)}
Y_{K{\ell}{m}}(\chi,\theta ,\varphi),
\label{S_Y_idnt}
\end{equation}
allows to cast the decompositions in the following matrix form,
\begin{table}[H]\label{chi_functions}
\footnotesize{
\begin{center}
       \begin{tabular}{|c|c|c|l c p{8.25cm}|}
%----------------------------------------------------------------------------------------------------------------------------
\hline
&&&&& \\
$K$ & $\widetilde{\ell}$ & $\widetilde{\ell} \leq \ell \leq K$ & $\psi^{\widetilde{\ell}}_{K} (\chi) $
& = &$ \sum_{\ell = \widetilde{\ell}}^{K} C_{\ell}\, \mathcal{S}^{\ell}_{K} (\chi) $ \\[3mm]
\hline   \hline
&&&&& \\
0 & 0 & 0 & $\psi^{0}_{0} (\chi)$ & = &$ \mathcal{S}^{0}_{0} (\chi)$ \\[3mm]
\hline
&&&&& \\
1 & 0 & 0,\,1 & $\psi^{0}_{1} (\chi)$ & = &$\mathcal{S}^{0}_{1} (\chi)\, +\, b\, \mathcal{S}^{1}_{1} (\chi) $ \\[2.5mm]
1 & 1 & 1 & $\psi^{1}_{1} (\chi)$ & = &$ \mathcal{S}^{1}_{1} (\chi) $ \\[3mm]
\hline
&&&&& \\
2 & 0 & 0,\,1,\,2 & $\psi^{0}_{2} (\chi)$
& = &$\mathcal{S}^{0}_{2} (\chi) \,+\, b\, \mathcal{S}^{1}_{2} (\chi) \,+\, ({2b \over 3})^2\, \mathcal{S}^{2}_{2} (\chi) $ \\[3mm]
2 & 1 & 1,\,2 & $\psi^{1}_{2} (\chi)$
& = &$\mathcal{S}^{1}_{2} (\chi) \,+\, {2\over 3}\, b\, \mathcal{S}^{2}_{2} (\chi) $ \\[3mm]
2 & 2 & 2 & $\psi^{2}_{2} (\chi)$
&= &$ \mathcal{S}^{2}_{2} (\chi) $ \\[3mm]
\hline
&&&&& \\
3 & 0 & 0,\,1,\,2,\,3 & $\psi^{0}_{3} (\chi)$
& = &$\mathcal{S}^{0}_{3} (\chi) \,+\, {9 \over 10}\, b\, \mathcal{S}^{1}_{3} (\chi) $
 $ +\, {b^2 \over 2}\, \mathcal{S}^{2}_{3} (\chi) $
 $ +\, \left( \frac{b^3}{ 8}-\frac{2b}{5} \right)\, \mathcal{S}^{3}_{3} (\chi) $ \\[3mm]
3 & 1 & 1,\,2,\,3 & $\psi^{1}_{3} (\chi)$
& = &$\mathcal{S}^{1}_{3} (\chi) \,+\, {5 \over 6}\, b\, \mathcal{S}^{2}_{3} (\chi) $
 $ +\, ({b \over 2})^2\, \mathcal{S}^{3}_{3} (\chi) $ \\[3mm]
3 & 2 & 2,\,3 & $\psi^{2}_{3} (\chi)$
& = &$\mathcal{S}^{2}_{3} (\chi) \,+\, {b\over 2}\,  \mathcal{S}^{3}_{3} (\chi) $ \\[3mm]
3 & 3 & 3 & $\psi^{3}_{3} (\chi)$
& = &$ \mathcal{S}^{3}_{3} (\chi) $ \\[3mm]
\hline
       \end{tabular}
\end{center}
}
\caption{The $\psi_K^{\ell}(\chi)$ parts of the solutions of the perturbed motion
in eq.~(\ref{S1_6})
in the basis of the free motion in eq.~(\ref{S1_2}).}
\end{table}

\begin{eqnarray}
\left(\begin{array}{c}
\Psi_{100}(\chi,\theta ,\varphi)\\
\Psi_{11{\widetilde m}}(\chi, \theta, \varphi)
\end{array}
\right)&=&e^{-\frac{\alpha_1\chi }{2}}
{\mathbf A}_1(\theta, \varphi)
\left(
\begin{array}{c}
Y_{100}(\chi, \theta, \varphi)\\
Y_{111}(\chi, \theta, \varphi)
\end{array}
\right),
\label{sl1}\\
\left(
\begin{array}{c}
\Psi_{200}(\chi, \theta ,\varphi)\\
\Psi_{21{\widetilde m}_1}(\chi, \theta, \varphi)\\
\Psi_{22{\widetilde m}_2}(\chi, \theta, \varphi)
\end{array}
\right)&=&e^{-\frac{\alpha_2\chi}{2}}
{\mathbf A}_2(\theta, \varphi)
\left(
\begin{array}{c}
Y_{200}(\chi, \theta, \varphi)\\
Y_{211}(\chi,\theta,\varphi)\\
Y_{222}(\chi, \theta, \varphi)
\end{array}
\right),
\label{sl2}\\
\left(
\begin{array}{c}
\Psi_{300}(\chi, \theta ,\varphi)\\
\Psi_{31{\widetilde m}_1}(\chi, \theta, \varphi)\\
\Psi_{32{\widetilde m}_2}(\chi, \theta, \varphi)\\
\Psi_{33{\widetilde m}_3}(\chi, \theta, \varphi)
\end{array}
\right)&=&e^{-\frac{\alpha_3\chi }{2}}
{\mathbf A}_3(\theta, \varphi)
\left(
\begin{array}{c}
Y_{300}(\chi, \theta ,\varphi)\\
Y_{311}(\chi, \theta, \varphi)\\
Y_{322}(\chi, \theta, \varphi)\\
Y_{333}(\chi, \theta, \varphi)
\end{array}
\right),\label{sl3}
\end{eqnarray}
where we  chose $m=\ell$ in (\ref{S_Y_idnt}).
The matrices ${\mathbf A}_K(\theta, \varphi)$ operate on the space of
exponentially scaled (damped) pseudo-spherical harmonics,
\begin{eqnarray}
{\widetilde Y}_{K{\ell}{ m}}(\chi, \theta, \varphi)&=&
e^{-\frac{\alpha_K\chi }{2}}Y_{K{ \ell}{ m}}(\chi, \theta, \varphi).
\label{Y_damped}
\end{eqnarray}
The explicit expressions for the ${\mathbf A}_K(\theta,\varphi)$
matrices for the  lowest $K$ values are given by,
\begin{eqnarray}
{\mathbf A}_1(\theta,\varphi)&=&
\left(
\begin{array}{cc}
1&\frac{b e^{-i\varphi}}{P_1^1}\\
0&
\frac{e^{i({\widetilde m}-1)\varphi}P_1^{{\widetilde m}}}{P_1^1}
\end{array}
\right),
\label{Ma1}\\
{\mathbf A}_2(\theta,\varphi)&=&
\left(
\begin{array}{ccc}
1&b\frac{e^{-i\varphi}}{P_1^1}&\frac{(2b)^2}{3^2}
\frac{e^{-2i\varphi}}{P_2^2}
\\
0&\frac{e^{i({\widetilde m}_1-1)\varphi}P_1^{{\widetilde m}_1}}{P_1^1}&
\frac{2b}{3}\frac{e^{i({\widetilde m}_1-2)\varphi}P_1^{{\widetilde m}_1}}{P_2^2}\\
0&0&\frac{e^{i({\widetilde m}_2-2)\varphi}P_2^{{\widetilde m}_2}}{P_2^2}
\end{array}
\right),
\label{Ma2}\\
{\mathbf A}_3(\theta,\varphi)&=&
\left(
\begin{array}{cccc}
1&\frac{9be^{-i\varphi}}{10P_1^1}&\frac{b^2e^{-i2\varphi}}{2P_2^2}&
\left( \frac{b^3}{8}-\frac{2b}{5}\right)\frac{e^{-i3\varphi }}{P_3^3}\\
0&\frac{e^{i({\widetilde m}_1-1)\varphi}P_1^{{\widetilde m}_1}}{P_1^1}&
\frac{5be^{i({\widetilde m}_1-2)\varphi }P_1^{{\widetilde m}_1}}{6P_2^2}&
\frac{b^2e^{i({\widetilde m}_1-3)\varphi}
P_1^{{\widetilde m}_1}}
{2^2P_3^3}\\
0&0&\frac{e^{i({\widetilde m}_2-2)\varphi }P_2^{{\widetilde m}_2}}{P_2^2}&
\frac{be^{i({\widetilde m}_2-3)\varphi}P_2^{{\widetilde m}_2}}{2P_3^3}\\
0&0&0&\frac{e^{i({\widetilde m}_3-3)\varphi}P_3^{{\widetilde m}_3}}{P_3^3}
\end{array}
\right),
\label{Ma3}
\end{eqnarray}
where we have dropped the $\cos \theta $ argument of the associated
Legendre functions, $P_{\ell}^{\ell}(\cos\theta)$,  
for the sake of simplifying the notations.
In noticing that $P_{\ell}^{\ell}(\cos\theta)\sim \sin^{\ell} \theta$, we 
observe that the
matrices connecting the representation functions of the perturbed and
free motions on $S^3$ are invertable in the entire open interval,
$\theta \in (0,\pi)$, being singular only at the poles.
Invertable matrices ${\mathbf A}_K(\theta,\varphi)$
at the poles can be obtained by replacing 
$P_l^l(\cos\theta)$ in eqs.~(\ref{Ma1})--(\ref{Ma3}),
through either $P_l^0(\cos 0)$ (``North'' pole), or $P_l^0(\cos \pi) $ 
(``South'' pole),  this because the special values of the
Legendre polynomials are finite at 
the ends of the interval under consideration. 
On $S^2$ the related matrices depend on the azimuthal angle 
$\varphi$ alone and are completely  free from singularities.

\subsection{Scaling similarity transformation of the
geometric so(4) algebra on $S^3$ to the algebra of 
the perturbed motion}

\begin{figure}
{\includegraphics[width=7.63cm]{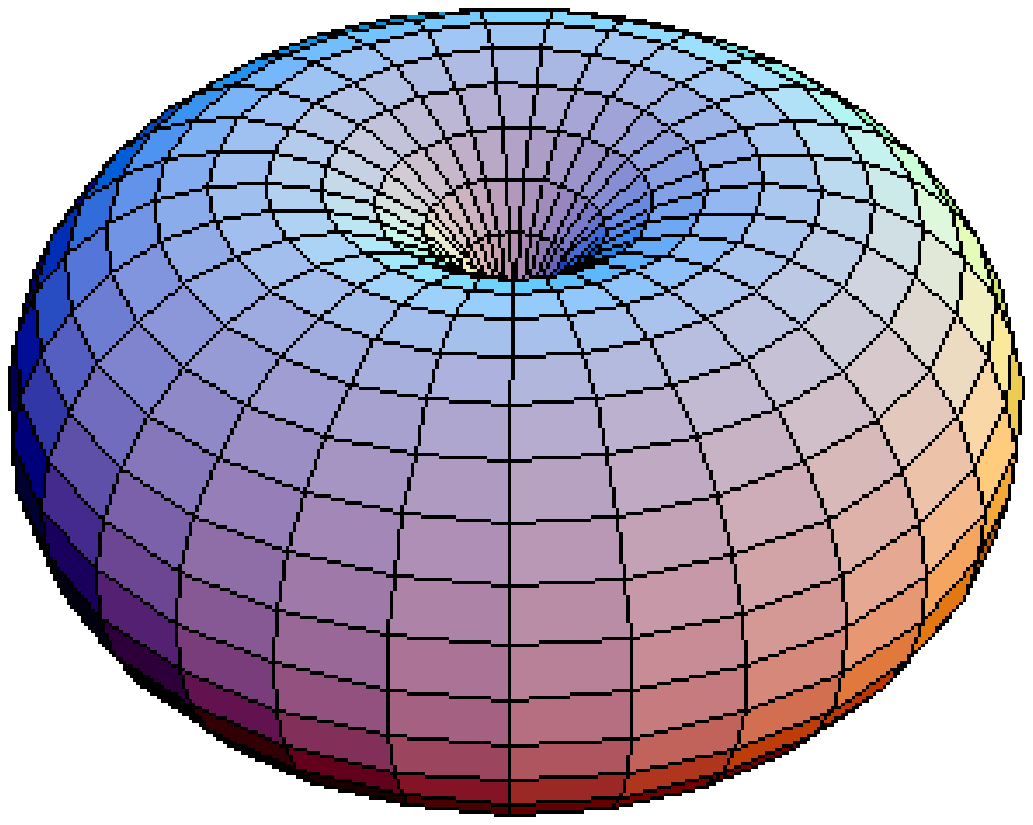}}
{\includegraphics[width=6.25cm]{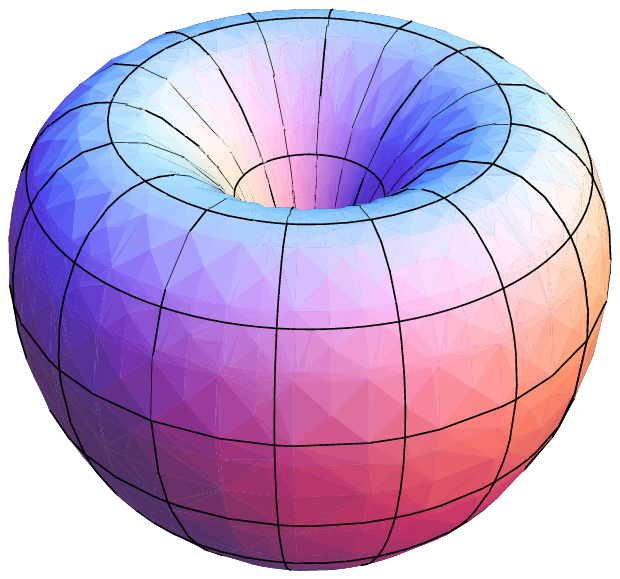}}
\caption{
The regular spherical harmonic $|Y_{111}(\chi, 0 , \varphi)|$ (left)
versus the damped, $|{\widetilde Y}_{111}(\chi,0, \varphi)|$, (right)
for b=2.
\label{S11}}
\end{figure}

\begin{figure}
{\includegraphics[width=5.9cm]{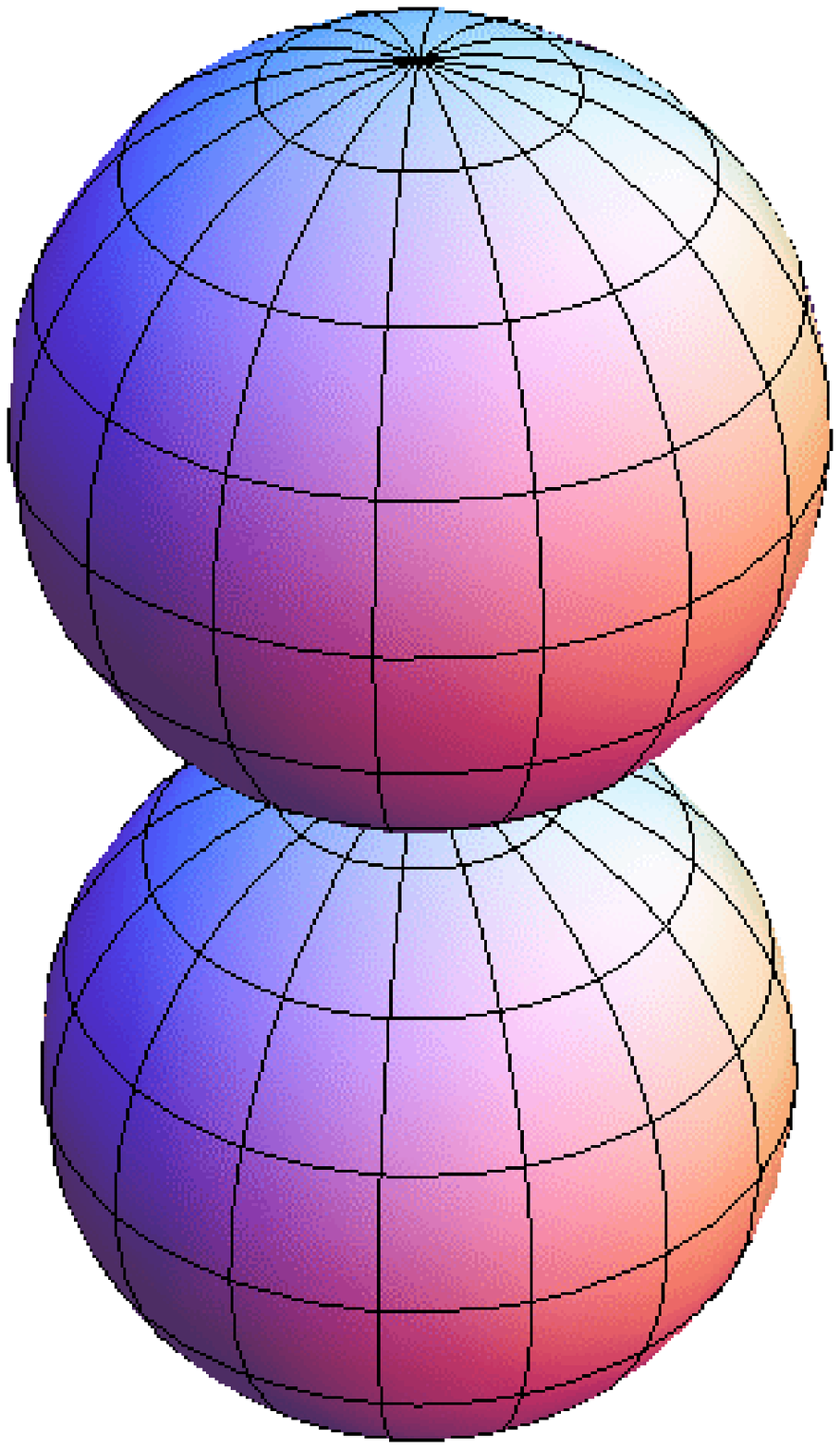}}
{\includegraphics[width=5.9cm]{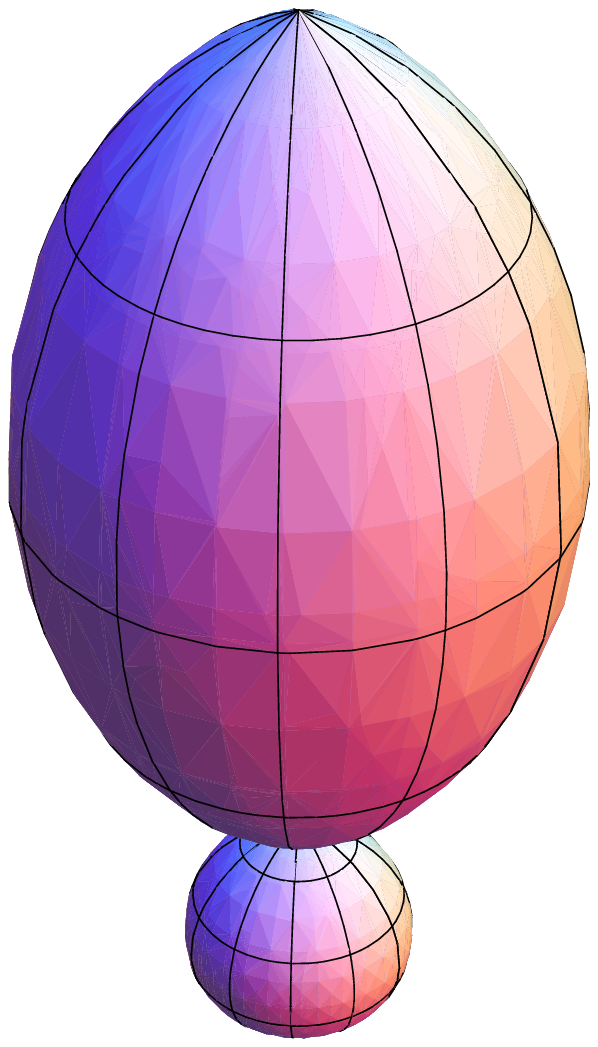}
%{Ydampedn_1m_0.eps}
}
\caption{
The regular hyper-spherical harmonic $|Y_{100}(\chi,0 , \varphi)|$ (left)
versus the damped, $|{\widetilde Y}_{100}(\chi,0, \varphi)|$, (right) for 
b=0.45.
\label{S10}}
\end{figure}

Substituting (\ref{S1_8}) in (\ref{I1_1}) and dragging
the exponential factor from the
very right to the very left results in

\begin{eqnarray}
e^{-\frac{\alpha_K \chi}{ 2}}
\left(
-{1 \over \sin^2 \chi }{\partial \over \partial \chi} \sin^2 \chi
{\partial \over \partial \chi}
+{\widetilde{\ell}(\widetilde{\ell} +1) \over \sin^2 \chi}  -
{\alpha_K^2 \over 4}  +\alpha_K\pmb D_K
\right)
\sum_{\ell={\widetilde \ell}}^K C_{\ell}{\mathcal S}_K^{\ell}(\chi)
&&\nonumber\\
= \left(-\frac{\alpha_K^2}{4} +K(K+2) \right) \,
\sum_{\ell={\widetilde \ell}}^K C_{\ell} e^{-\frac{\alpha_K \chi}{ 2}}{\mathcal S}_K^{\ell}(\chi).&&
\label{R_1}
\end{eqnarray}
Here, we introduced the differential operator $\pmb D_K$ as,
\begin{eqnarray}\label{R_2}
\pmb D_K &=&\left( {\partial \over \partial \chi} -K \cot \chi \right),
\end{eqnarray}
and  used  $\alpha_K = 2b/(K+1)$ .
For ${\widetilde \ell}$ taking the maximal value,
${\widetilde \ell}=K$, the operator $\pmb D_K$ coincides with the
raising operator of the so(4) algebra and  nullifies
${\mathcal S}_K^K(\chi)\sim \sin^K\chi$.
In effect, one encounters the identity,
\begin{eqnarray}
\left[{\mathcal K} -2b\cot\chi \right]
\Psi_{KK{\widetilde m}}(\chi, \theta, \varphi)&=&
\left[ e^{-\frac{\alpha_K \chi}{2}}\,
\left( {\mathcal K} - {\alpha_K^2 \over 4}
\right)e^{\frac{\alpha_K \chi}{ 2}}\right]
\Psi_{KK{\widetilde m}}(\chi,\theta, \varphi)\nonumber\\
&=& \left( K(K+2)-\frac{\alpha_K^2}{4}  \right) \,
\Psi_{KK{\widetilde m}}(\chi,\theta, \varphi).
\label{caso1}
\end{eqnarray}
The latter expression  shows that
up  to a non-unitary scaling transformation, and a shift by a constant,
the eigenvalue problem of the
cotangent perturbed geodesic motion on $S^3$ for $|KK{\widetilde m}>$
results equivalent to the
eigenvalue problem of the free geodesic motion.
Such in fact is valid for any set of the quantum numbers,
$ |K{\widetilde \ell}{\widetilde m}>$,
this by virtue of the following recurrence relations among 
${\mathcal S}_K^l(\chi)$ functions, which translate into recurrence 
relations among  Gegenbauer polynomials,
\begin{eqnarray}
\pmb{D}_{1} \,\mathcal{S}^{0}_{1} (\chi)\,
=2 \, \csc^2 \chi \, \mathcal{S}^{1}_{1} (\chi),
&\quad&
\pmb{D}_{2} \,\mathcal{S}^{1}_{2} (\chi)= 4 \, \csc^2 \chi \,
\mathcal{S}^{2}_{2} (\chi),
\nonumber\\
\pmb{D}_{2} \,\mathcal{S}^{0}_{2} (\chi)\,
= 2 \, \csc^2 \chi \, \mathcal{S}^{1}_{2} (\chi),&\quad&
\pmb D_3\, \mathcal{S}_3^2(\chi)=6\csc^2\chi\,{\mathcal S}_3^3(\chi),  
\nonumber\\
&\quad &\pmb D_3\,{\mathcal S}_3^1(\chi)=\frac{20}{3}\csc^2\chi{\mathcal S}_3^2(\chi),
\nonumber\\
\pmb{D}_{K}\,\mathcal{S}_K^{K} (\chi) = 0,\quad \forall K, \,\,
\mbox{etc.},&\quad&
{\mathcal S}_K^{\ell}(\chi)=\sin^{\ell}\chi {\mathcal G}_{K-\ell}^{\ell+1}(\cos \chi).
\label{RRR}
\end{eqnarray}
In order to illustrate the r\'ole of the recurrence relations we here
present the simple example  of $\Psi_{100}(\chi, \theta, \varphi)$.
In this case, and according to Table 1, equation (\ref{S1_8}) reduces to
\begin{eqnarray}
\psi_1^0(\chi)&=&{\mathcal S}_1^0(\chi)  +b{\mathcal S}_1^1(\chi),\nonumber\\
{\mathcal S}_1^0(\chi) =-2\sin\chi \cot \chi,
&\quad &{\mathcal S}_1^1(\chi) =\sin\chi.
\label{Exmpl_1}
\end{eqnarray}
Upon substitution of (\ref{Exmpl_1}) in (\ref{R_1}),
it is straightforward showing that by virtue of the first relation in
(\ref{RRR}), the $\alpha_1\pmb D_1{\mathcal S}_1^0(\chi)$ term produces the
exact centrifugal term of the second component, $b{\mathcal S}_1^1(\chi)$,
so that the net action of $({\mathcal K}-2b\cot \chi)$
on  $e^{-\frac{\alpha_1\chi}{2}}\psi_{1}^0(\chi)$ becomes,
\begin{eqnarray}
\bigg(
{\mathcal K}-2b\cot\chi \bigg)
e^{-\frac{\alpha_1\chi }{2}}
\bigg( {\mathcal S}_1^0(\chi)+b{\mathcal S}_1^1(\chi)\bigg)&=&
\left[ e^{-\frac{\alpha_1\chi }{2}}
\left( {\mathcal K}-\frac{\alpha_1^2}{4}
\right) e^{\frac{\alpha_1\chi }{2}}\right]
\left[ e^{-\frac{\alpha_1\chi }{2}}
\bigg( {\mathcal S}_1^0(\chi)+b{\mathcal S}_1^1(\chi)\bigg)\right]
\nonumber\\
&=&\left(3 -
\frac{\alpha_1^2}{4}
\right) e^{-\frac{\alpha_1\chi}{2}}
\psi_{1}^0(\chi).
\label{bong}
\end{eqnarray}
In Figs.~\ref{S11}, and \ref{S10}, the modification of the
shapes of hyper-spherical and damped hyper-spherical harmonics 
corresponding to the case in eq.~(\ref{Exmpl_1}),
are shown for illustrative purposes.

Therefore, at the general level of the total wave functions, one finds,
\begin{eqnarray}
\left[{\mathcal K} -2b\cot\chi \right]
\Psi_{100}(\chi, \theta, \varphi)&=&
\left[ e^{-\frac{\alpha_K \chi}{2}}\,
\left( {\mathcal K} - {\alpha_K^2 \over 4}
\right)e^{\frac{\alpha_K \chi}{ 2}}\right]
\Psi_{100}(\chi,\theta, \varphi)\nonumber\\
&=& \left( K(K+2)-\frac{\alpha_K^2}{4}  \right) \,
\Psi_{100}(\chi,\theta, \varphi).
\label{caso2}
\end{eqnarray}

In fact, the recurrence relations in (\ref{RRR}) guarantee validity of,
\begin{eqnarray}
\left( \frac{{\widetilde \ell}({\widetilde \ell}+1)}{\sin^2\chi}
+\alpha_K{\mathbf D}_K\right)\sum_{\ell={\widetilde \ell}}^{\ell=K}C_{\ell}\,
{\mathcal S}_K^{\ell}
&=&
\sum_{\ell={\widetilde \ell}}^{\ell=K}\frac{\ell(\ell+1)}{\sin^2\chi}C_{\ell}\,{\mathcal S}_K^{\ell},
\label{bing_bong}
\end{eqnarray}
which, upon substitution in (\ref{R_1}), allows to generalize  
the transformation in (\ref{caso1}) and (\ref{caso2})
to any $\Psi_{K{\widetilde \ell}{\widetilde m}}(\chi,\theta, \varphi)$.
In consequence, the similarity transformation between the
$({\mathcal K}-2b\cot\chi)$ and the ${\mathcal K}$ eigenvalue problems
can be cast in the following matrix form,

\begin{eqnarray}
\left( {\mathcal K}-2b\cot\chi\right)
{\mathbf X}_K(\chi,\theta,\varphi)
&=&
\left[ e^{-\frac{\alpha_K\chi }{2}}{\mathbf A}_K(\theta,\varphi)
\left({\mathcal K}- \frac{\alpha_K^2}{4} \right)
e^{\frac{\alpha_K\chi }{2}}{\mathbf A}^{-1}_K(\theta,\varphi)\right]\, 
{\mathbf X}_K(\chi,\theta,\varphi)\nonumber\\
{\mathbf X}_K(\chi,\theta,\varphi)&=&\left(
\begin{array}{c}
\Psi_{K00}(\chi,\theta,\varphi)\\
\Psi_{K1{\widetilde m}_1}(\chi,\theta,\varphi)\\
...\\
\Psi_{KK{\widetilde m}_K}(\chi,\theta,\varphi)
\end{array}
\right).
\label{voala}
\end{eqnarray}
The latter equation shows that the eigenvalue problem
of the cotangent perturbed motion on $S^3$ can be represented
as the eigenvalue problem of a similarity-transformed  
Casimir operator, ${\mathcal K}$, of
the ordinary so(4) isometry  algebra, 
shifted by a constant (within the representation space of interest).
 The transformation is non-unitary
and of the dilation type. 
As long as the similarity transformation of the  Casimir invariant
of the geometric algebra can be viewed as the result of subjecting the
so(4) elements  $L_i$, and $N_j$ in (\ref{S1_3}) to same transformation,
we here have designed a representation of the geometric so(4) algebra
for a ``curved'' Coulomb potential  on $S^3$.

\begin{figure}[htb]
\vskip 3.1cm
\includegraphics{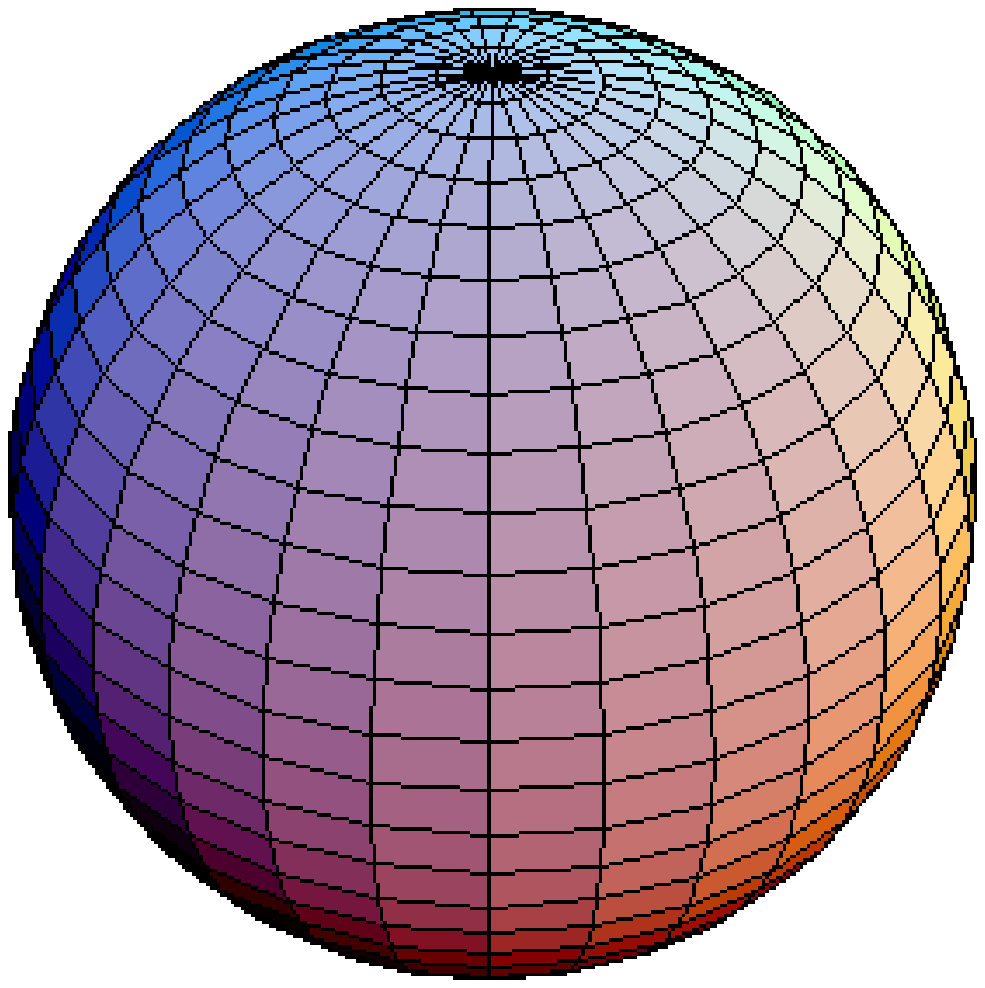}
\includegraphics{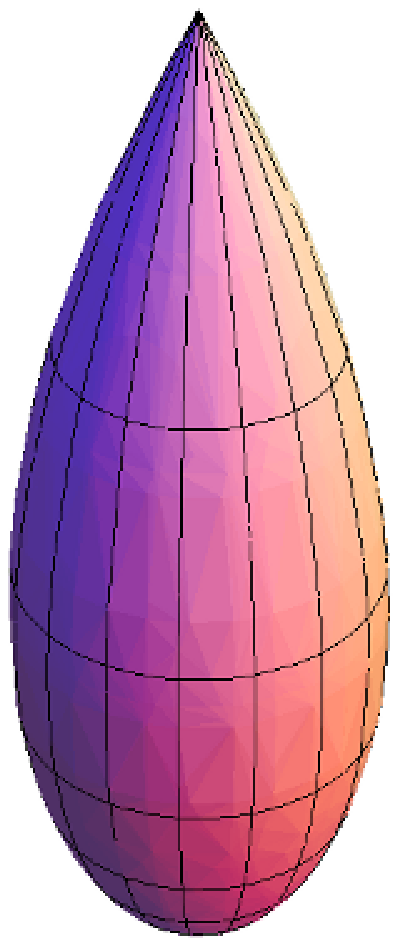}
\vspace{0.95cm}
{\caption{ Deformation of the spherical metric, $|Y_{000}(\chi, 0\,\varphi)|$,
(left) versus the exponentially scaled one, 
$|\widetilde {Y}_{000}(\chi, 0,\varphi )|$ (right) for b=1.}
}
\label{mtrc}
\end{figure}

\section{Conclusions}
The clue of the present study is that modulo a shift by a constant,   
a non-unitary scaling similarity transformation converts
the Casimir invariant of the so(4) isometry algebra of the three--ball $S^3$
into the Coulomb potential problem there.
The transformation, presented in (\ref{voala}),
was concluded from decomposing the eigenfunctions of the
``curved'' Coulomb potential problem
in the bases of exponentially scaled (damped) hyper-spherical harmonics
in (\ref{sl1})--(\ref{Y_damped}).
It was furthermore shown to emerge by virtue  of specific
recurrence relations (\ref{RRR}) among Gegenbauer polynomials.

The particle motion on $S^3$ considered here
had the peculiarity that the  perturbance left  the degeneracy
patterns characterizing  the free motion intact, though  the
isometry group symmetry has been broken. The breaking of the initial
SO(4) rotational invariance by the ``curved'' Coulomb potential happened
at the level of the representation functions, thus avoiding
the more severe  breakdown at the level of
the algebra through deformations \cite{Higgs}, \cite{Leemon},
\cite{Quesne}, \cite{AsimG}.
This subtle type of symmetry breaking is visualized in
Fig.~4 by the
deformation of the metric of the $S^3$ sphere  through the
damping exponential  factor under consideration.
It might be interesting to notice,
that such a  metric deformation  emerges also in effect of a
conformal symmetry  breaking by the dilaton mass \cite{Blaschke}.
The method proposed can easily  be extended towards  hyperbolic spaces
through complexification of the polar angle $\chi$ in (\ref{I2}) 
as $\chi \to i\chi $. Simultaneously changing $b\to -ib$ in (\ref{I6})
takes one to the Coulomb problem on a hyperboloid.
Quantum problems on hyperbolic spaces, pioneered by Dirac 
\cite{Dirac35} through his consideration of the electron 
wave equation in the De-Sitter universe,
acquire  importance as well within the context of pure relativistic 
descriptions of 
bound systems, as within the context of gravitational studies.

\section{Acknowledgments}

We thank Dr. Jose Antonio Vallejo for an illuminating discussion on
the relationship between groups and algebras.
Work partly supported by CONACyT-M\'{e}xico under grant number
CB-2006-01/61286.


\begin{thebibliography}{99}

\bibitem{PRA} Jean-Claude Gay, Dominique Delande, and Antoine Bommier,
{\it Atomic quantum states with maximal localization on classical
elliptical orbits,\/} Phys.\ Rev.\ A {\bf 39}, 6587--6590 (1989).

\bibitem{PRE} Reimar Finken, Matthias Schmidt, and Hartmut L\"owen,
{\it Freezing transition of hard hyperspheres,\/}
Phys.\ Rev. E {\bf 65}, 016108 (2001).

\bibitem{PRL} Pierre-Francois Loos, Peter M. W. Gill,
{\it Two electrons on a Hypersphere;A Quasiexactly Solvable Model},
Phys.\ Rev.\ Lett.\ {\bf 103}, 123008 (2009).

\bibitem{poly} Per Johan R\'asmark, Tobias Ekholm, and Christer Elvingson,
{\it Computer simulations of polymer chain structrure and dynamics ona a hypersphere in four-space,\/} J.\ Chem.\ Phys.\ {\bf 122}, 184110 (2005).


\bibitem{Brown}
Jarl Nissfolk, Tobias Ekholm, and Christer Elvingson,
{\it Brownian dynamics simulations on a hypersphere in 4-space,\/}
J.\ Chem.\ Phys.\ {\bf 119} 6423--6432 (2003).


\bibitem{qtm_dots} V.\ V.\ Gritsev, Yu.\ A.\ Kurochkin,
{\it Model of excitations in quantum dots based on quantum mechanics
in spaces of constant curvature,\/}
Phys.\ Rev.\ B {\bf 64}, 035308-1-035308-9 (2001).

\bibitem{Higgs}P.\ W.\  Higgs,
{\it Dynamical symmetries in a spherical geometry,\/}
J.\ Phys A:Math.Gen. {\bf 12}, 309-323 (1979).

\bibitem{MolPhys} D.\ E.\ Alvarez-Castillo, C.\ B.\ Compean, and M.\ Kirchbach,
{\it Rotational symmetry and degeneracy: a cotangent perturbed rigid
rotator of unperturbed level multiplicity,\/} Mol.\ Phys.\ {\bf 109}, 1477-1483
(2011).

\bibitem{Leemon} Howard Leemon, {\it Dynamical symmetries in a
spherical geometry,\/} J.\ Phys.\ A:Math.Gen. {\bf 14}(4), 489--501 (1979).


\bibitem{Quesne} C.\ Quesne, V.\ M.\ Tkachuk,
{\it Deformed algebras, position dependent effective masses and curved space:
An exactly solvable Coulomb problem,\/} J.\ Phys.\ A {\bf 37},
4267--4281 (2004).




\bibitem{raposo} A.\ Raposo, H.-J.\ Weber, D.\ E.\ Alvarez-Castillo, and
M.\ Kirchbach, {\it Romanovski polynomials in selected physics problems,\/}
C.\ Eur.\ J.\ Phys.\ {\bf 5}, 253-284 (2007).

\bibitem{Engelfield} M.\ J.\ Englefield,
{\it Group Theory and the Coulomb Problem\/} (Wiley-Interscience, N.Y.,1971).

\bibitem{Dutt} R.\ De, R.\ Dutt, U.\ Sukhatme,
{\it Mapping of shape invariant potentials under point
canonical transformations, \/} J.\ Phys.\ A:Math.Gen. {\bf 25},
L843-L850 (1992).

\bibitem{Schr40} E.\ Schr\"odinger,
{\it A method of determining quantum mechanical eigenvalues and
eigenfunction,\/}
Proc.\ Roy.\ Irish Acad.\ A {\bf 46}, 9-16 (1940).

\bibitem{Schr41} E.\ Schr\"odinger,
{\it Further studies on solving eigenvalue problems by factorization\/},
Proc.\ Roy.\ Irish Acad.\, {\bf 46}, 183-206 (1941).\\
http:://www.jstor.org/pss/20490756
 
\bibitem{Barut} A.\ O.\ Barut, Raj Wilson,
{\it On the dynamical group of the Kepler problem
in a curved space of constant curvature, \/}
Phys.\ Lett.\ A {\bf 110}, 351-354 (1985).



\bibitem{KC_10} M.\ Kirchbach, C.\ B.\ Compean,
{\it  Conformal symmetry and light flavor baryon spectra,\/}
Phys.\ Rev.\ D  {\bf 82}, 034008 (2010).

\bibitem{AsimG} Asim Gangopadhyaya, Jeffrey V.\ Mallow, and Uday P.\ Sukhatme,
{\it Translational shape invariance and the inherent potential algebra,\/}
Phys.\ Rev.\ A {\bf 58} 4287-4292 (1998).

\bibitem{Blaschke}M.\ P.\ Dabrowski, J.\ Garecki, and D.\ B.\ Blaschke,
{\it Conformal transformations and conformal invariance in gravity, \/}
Ann.\ d.\ Physik {\bf 18}, 13-22 (2009).


\bibitem{Dirac35} P.\ A.\ M.\ Dirac,
{\it The electron wave equation in De-Sitter space}, Annals of Math.\ Phys.\
{\bf 36}, 657-669 (1935).\\
http://www.jstor.org/stable/1968649


\end{thebibliography}
\end{document}